\documentclass[lnicst,sechang,a4paper]{svmultln}
\usepackage{amssymb}
\usepackage{graphicx}

\usepackage{url}
\urldef{\mailsa}\path|{parakh, subhashk}@cs.okstate.edu|
\usepackage{amsfonts}

\begin{document}

\mainmatter  

\title{A Distributed Data Storage Scheme for Sensor Networks}

\author{Abhishek Parakh \and Subhash Kak}
\institute{Computer Science Department\\
Oklahoma State University, Stillwater OK 74075\\
\mailsa}

\maketitle

\begin{abstract}
We present a data storage scheme for sensor networks that achieves the targets of encryption and distributed storage simultaneously. We partition the data to be stored into numerous pieces such that at least a specific number of them have to be brought together to recreate the data. The procedure for creation of partitions does not use any encryption key and the pieces are implicitly secure. These pieces are then distributed over random sensors for storage. Capture or malfunction of one or more (less than a threshold number of sensors) does not compromise the data. The scheme provides protection against compromise of data in specific sensors due to physical capture or malfunction.
\end{abstract}

\section{Introduction}
Sensors may deployed in hostile environment where they may be prone to dangers ranging from environmental hazards to physical capture. Under such circumstances the data and encryption keys stored on these sensors is vulnerable to compromise.

A number of techniques have been proposed for secure communication between sensors using key management and data encryption \cite{ref1, ref2, ref3, ref4}. Some techniques aim at secure routing \cite{ref5, ref6}, intrusion detection \cite{ref7} and others describe a mechanism for moving sensitive data around in the network from time to time \cite{ref8}. A few researchers propose distributed data storage \cite{ref15, ref16, ref17} but do not adequately address the question of security or use explicit encryption techniques to secure data which leaves the question of secure storage of encryption/decryption keys unanswered.

To go beyond the present approaches, one may incorporate further security within the system by using another layer that increases the space that the intruder must search in order to break a cipher \cite{ref9, ref10}. Here, we propose an implicitly secure data partitioning scheme whose security is distributed amongst many sensors. In contrast to hash-based distributed security models for wireless sensor networks \cite{ref18, ref19}, we consider a more general method of data partitioning. In this approach, stored data is partitioned into two or more pieces and stored at randomly chosen sensors on the network. In scenarios where one or more pieces may be at the danger of being lost or inaccessible due to sensor failure or capture, one may employ schemes that can recreate the data from a subset of original pieces.

A number of schemes have been proposed in the communications context for splitting and sharing of decryption keys \cite{ref11, ref20}. These schemes fall under the category of ``secret sharing schemes'', where the decryption key is considered to be a secret. Motivated by the need to have an analog of the case where several officers must simultaneously use their keys before a bank vault or a safe deposit box can be opened, these schemes do not consider the requirement of data protection for a single party. Further, in any secret sharing scheme it is assumed that the encrypted data is stored in a secure place and that none of it can be compromised without the decryption key.

In this paper, we protect data by distributing its parts over various sensors. The idea of making these partitions is a generalization of the use of 3 or 9 roots of a number in a cubic transformation \cite{ref12}. The scheme we present is simple and easily implementable.

We would like to stress that the presented scheme is different from Shamir's secret sharing scheme which takes the advantage of polynomial interpolation. Further, Shamir's scheme maps the secret as points on the y-axis,, whereas the scheme proposed in this paper maps the secret as points on the x-axis, as roots of a polynomial.

\section{Proposed Data Partitioning Scheme}

By the fundamental theorem of algebra, every equation of $k^{th}$ degree has $k$ roots. We use this fact to partition data into $k$ partitions such that each of the partition is stored on a different sensor. No explicit encryption of data is required to secure each partition. The partitions in themselves do not reveal any information and hence are implicitly secure. Only when all the partitions are brought together is the data revealed.

Consider an equation of degree $k$
\begin{equation}\label{eq1}
    x^k+a_{k-1}x^{k-1}+a_{k-2}x^{k-2}+...+a_1x+a_0=0
\end{equation}
Equation \ref{eq1} has $k$ roots denoted by \{$r_1, r_2,..., r_k$\} $\subseteq$ \{set of complex numbers\} and can be rewritten as
\begin{equation}\label{eq2}
    (x-r_1)(x-r_2)...(x-r_k)=0
\end{equation}
In cryptography, it is more convenient to use the finite field $\mathbb{Z}_p$ where $p$ is a large prime. If we replace $a_0$ in (\ref{eq1}) with the data $d\in\mathbb{Z}_p$ that we wish to partition then,
\begin{equation}\label{eq3}
    x^k+\sum_{i=1}^{k-1}a_{k-i}x^{k-i}+d\equiv 0 \bmod p
\end{equation}
where $0\leq a_i \leq p-1$ and $0\leq d \leq p-1$. (Note that one may alternatively use $-d$ in (\ref{eq3}) instead of $d$.) This may be rewritten as
\begin{equation}\label{eq4}
    \prod_{i=1}^{k}(x-r_i)\equiv 0 \bmod p
\end{equation}
where $1\leq r_i \leq p-1$. The roots, $r_i$, are the partitions. It is clear that the term $d$ in (\ref{eq3}) is independent of variable $x$ and therefore
\begin{equation}\label{eq5}
    \prod_{i=1}^{k}r_i\equiv d \bmod p
\end{equation}
If we allow the coefficients in (\ref{eq3}) to take values $a_1=a_2=...=a_{k-1}=0$, then (\ref{eq3}) will have $k$ roots only if $GCD({p-1},k) \neq 1$ and $\exists b\in \mathbb{Z}_p$ such that $d$ is the $k^{th}$ power of $b$. One simple way to chose such a $p$ would be to choose a prime of the form $(k\cdot s+1)$, where $s\in\mathbb{N}$. However, such a choice would not provide good security because knowledge of the number of roots and one of the partitions would be sufficient to recreate the original data by computing the $k^{th}$ power of that partition. Furthermore, not all values of $d$ will have a $k^{th}$ root and hence one cannot use any arbitrary integer, which would typically be required. Therefore, one of the restrictions on choosing the coefficients is that not all of them are simultaneously zero.

For example, if the data needs to be divided into two parts then an equation of second degree is chosen and the roots computed. If we represent this general equation by
\begin{equation}\label{eq6}
    x^2+a_1x+d\equiv 0 \bmod p
\end{equation}
then the two roots can be calculated by solving the following equation modulo $p$,
\begin{equation}\label{eq7}
    x=\frac{-a_1\pm\sqrt{a_{1}^2-4d}}{2}
\end{equation}
which has a solution in $\mathbb{Z}_p$ only if the square root $\sqrt{a_{1}^2-4d}$ exists modulo $p$. If the square root does not exist then a different value of $a_1$ needs to be chosen. We present a practical way of choosing the coefficients below. However, this brings out the second restriction on the coefficients, i.e. they should be so chosen such that a solution to the equation exists in $\mathbb{Z}_p$.

\begin{theorem}\label{theorem1}
If the coefficients $a_i$, $1\leq i\leq k-1$ in equation (\ref{eq3}) are not all simultaneously zero, are chosen randomly and uniformly from the field, then the knowledge of any $k-1$ roots of the equation, such that equation (\ref{eq4}) holds, does not provide any information about the value of $d$ with a probability greater than that of a random guess of $1/p$.
\end{theorem}
\begin{proof}
Given a specific $d$, the coefficients in (\ref{eq3}) can be chosen to satisfy (\ref{eq4}) in $\mathbb{Z}_p$ in the following manner. Choose at randomly and uniformly from the field $k-1$ random roots $r_1, r_2, ..., r_{k-1}$. Then $k^{th}$ root $r_k$ can be computed by solving the following equation,
\begin{equation}\label{eq8}
    r_k=d\cdot (r_1\cdot r_2\cdot ...\cdot r_{k-1})^{-1} \bmod p
\end{equation}
Since the roots are randomly chosen from a uniform distribution in $\mathbb{Z}_p$, the probability of guessing $r_k$ without knowing the value of $d$ is $1/p$. Conversely, $d$ cannot be estimated with a probability greater than $1/p$ without knowing the $k^{th}$ root $r_k$.\qed
\end{proof}

It follows from Theorem \ref{theorem1} that data is represented as a multiple of $k$ numbers in the finite field.

\textit{Example 1.}  Let data $d=10$, prime $p=31$, and let $k=3$. We need to partition the data into three parts for which we will need to use a cubic equation, $x^3+a_2x^2+a_1x-d\equiv 0\bmod p$.

We can find the equation satisfying the required properties using Theorem \ref{theorem1}. Assume, $(x-r_1)(x-r_2)(x-r_3)\equiv 0 \bmod 31$. We randomly choose 2 roots from the field, $r_1=19$ and $r_2=22$. Therefore, $r_3\equiv d\cdot (r_1\cdot r_2)^{-1}\equiv 10\cdot (19\cdot 22)^{-1}\bmod 31\equiv 11$.
The equation becomes $(x-r_1)(x-r_2)(x-r_3)\equiv x^3-21x^2+x-10\equiv 0\bmod 31$, where the coefficients are $a_1=1$ and $a_2=-21$ and the partitions are 11, 22 and 19.

{\textbf{Choosing the Coefficients:} We described above two conditions that must be satisfied by the coefficients. The first condition was that not all the coefficients are simultaneously zero and second that the choice of coefficients should result in an equation with roots in $\mathbb{Z}_p$. Since no generalized method for solving equations of degree higher than 4 exists \cite{ref21}, a numerical method must be used which becomes impractical as the number of partitions grows. An easier method to compute the coefficients is exemplified by Theorem \ref{theorem1} and Example 1.

One might ask why should we want to compute the coefficients if we already have all the roots? We answer this question a little later.

{\textbf{Introducing Redundancy:} In situations when the data pieces stored on one or more (less than a threshold number of) sensors over the sensor network may not be accessible, then other sensors should be able to collaborate to recreate the data from the available pieces. The procedure outlined below extends the $k$ partitions to $n$ partitions such that only $k$ of them need to brought together to recreate the data.
If $\{r_1, r_2, \ldots, r_k\}$ is the original set of partitions then they can be mapped into a set of $n$ partitions $\{p_1, p_2, \ldots, p_n\}$ by the use of a mapping function based on linear algebra. If we construct $n$ linearly independent equations such that
\[\begin{array}{c}
a_{11}r_1+a_{12}r_2+\ldots +a_{1k}r_k=c_1\\
a_{21}r_1+a_{22}r_2+\ldots +a_{2k}r_k=c_2\\
\vdots\\
a_{n1}r_1+a_{n2}r_2+\ldots +a_{nk}r_k=c_n
\end{array}\]
where numbers $a_{ij}$ are randomly and uniformly chosen from the finite field $\mathbb{Z}_p$, then the $n$ new partitions are $p_i=\{a_{i1}, a_{i2}, \ldots, a_{ik},\:c_i\}$, $1\leq i\leq n$.
The above linear equation can be written as matrix operation,
\[
\left[ \begin{array}{cccc}
a_{11}&a_{12}&\ldots&a_{1k}\\
a_{21}&a_{22}&\ldots&a_{2k}\\
\vdots&&&\\
a_{n1}&a_{n2}&\ldots&a_{nk} \end{array} \right]
\left[ \begin{array}{c}
r_1\\
r_2\\
\vdots\\
r_k \end{array} \right]=
\left[ \begin{array}{c}
c_1\\
c_2\\
\vdots\\
c_n \end{array} \right]
\]
To recreate $r_j$, $1\leq j\leq k$ from the new partitions, any $k$ of them can be brought together,
\[
\left[ \begin{array}{c}
r_1\\
r_2\\
\vdots\\
r_k \end{array} \right]=
\left[ \begin{array}{cccc}
a_{m1} & a_{m2} & \ldots & a_{mk}\\
a_{n1} & a_{n2} & \ldots & a_{nk}\\
\vdots&&&\\
a_{i1} & a_{i2} & \ldots & a_{ik} \end{array}
\right]_{k\times k}^{-1}
\left[ \begin{array}{c}
c_1\\
c_2\\
\vdots\\
c_k \end{array}
\right]_{k\times 1}
\]

A feature of the presented scheme is that new partitions may be added and deleted without affecting any of the existing partitions.

{\textbf{An alternate approach to partitioning:} Once a sensor has computed all the $k$ roots then it may compute the equation resulting from (\ref{eq4}) and store one or all of the roots on different sensors and the coefficients on different sensors. Recreation of original data can now be performed in two ways: either using (\ref{eq5}) or choosing one of the roots at random and retrieving the coefficients and substituting the appropriate values in the equation (\ref{eq3}) to compute
\begin{equation}\label{eq9}
    -a_{0}\equiv x^k+a_{k-1}x^{k-1}+a_{k-2}x^{k-2}+...+a_1x\bmod p
\end{equation}
Therefore, the recreated data $d\equiv (p-a_0)\bmod p$. Parenthetically, this may provide a scheme for fault tolerance and partition verification. Additionally, a sensor may store just one of the roots and $k-1$ coefficients on the network. These together represent $k$ partitions.

\textit{Note.} Distinct sets of coefficients (for a given constant term in the equation) result in distinct sets of roots and vice versa. This is because two distinct sets of coefficients represent distinct polynomials because two polynomials are said to be equal if and only if they have the same coefficients. By the fundamental theorem of algebra, every polynomial has unique set of roots.

Two sets of roots $R_1$ and $R_2$ are distinct if and only if $\exists r_i \forall r_j (r_i\neq r_j)$, where $r_i\in R_1$ and $r_j\in R_2$. To compute the corresponding polynomials and the two sets coefficients $C_1$ and $C_2$, we perform $\prod_{{i=1},{r_i\in R_1}}^{k}(x-r_i)\equiv 0\bmod p$ and $\prod_{{j=1},{r_j\in R_2}}^{k}(x-r_j)\equiv 0\bmod p$ and read the coefficients from the resulting polynomials, respectively. It is clear the at least one of the factors of the two polynomials is distinct because at least one of the roots is distinct; hence the resulting polynomials for a distinct set of roots are distinct.

\begin{theorem}\label{thm3}
Determining the coefficients of a polynomial of degree $k\geq 2$ in a finite field $\mathbb{Z}_p$, where $p$ is prime, by brute force, requires $\Omega(\lceil\frac{p^{k-1}}{(k-1)!}\rceil)$ computations.
\end{theorem}
\begin{proof}
If $A$ represents the set of coefficients \\$A=\{a_1, a_2, ..., a_{k-1}\}$, where $0\leq a_i \leq p-1$, then by the above note each distinct instance of set $A$ gives rise to a distinct set of roots $R=\{r_1, r_2, ..., r_k\}$, and, conversely, every distinct instance of set $R$ gives rise to a distinct set of coefficients $A$. Therefore, if we fix $d$ to a constant, then (\ref{eq5}) can be used to compute the set of roots. Every distinct set of roots is therefore a $k-1$ combination of a multiset \cite{ref13, ref14}, where each element has infinite multiplicity, and equivalently a $k-1$ combination of set $S=\{0,1,2, ...,p-1\}$ with repetition allowed. Thus, the number of possibilities for the choices of coefficients is given by the following expression
\begin{equation}\label{eq10}
\begin{array}{ll}
    \left< \begin{array}{cc}
        p\\
        k-1
    \end{array}\right>\vspace{.1in}
    &=\left( \begin{array}{cc}
        p-1+(k-1)\\
        k-1
    \end{array}\right)\\\vspace{.1in}
    &=\frac{(p+k-2)!}{(p-1)!(k-1)!}\\\vspace{.1in}
    &=\frac{(p+k-2)(p+k-3)...(p+k-k)(p-1)!}{(p-1)!(k-1)!}\\\vspace{.1in}
    &=\frac{(p+k-2)(p+k-3)...p}{(k-1)!}\\ \vspace{.1in}
    &\geq \lceil\frac{p^{k-1}}{(k-1)!}\rceil
\end{array}
\end{equation}
Here we have used the fact that in practice $p\gg k\geq2$, hence the result. We have ignored the one prohibited case of all coefficients being zero, which has no effect on our result.\qed \end{proof}

\section{A stronger variation to the protocol modulo a composite number}
An additional layer of security may be added to the implementation by performing computations modulo a composite number $n=p\cdot q$, $p$ and $q$ are primes, and using an encryption exponent to encrypt the data before computing the roots of the equation. In such a variation, knowledge of all the roots and coefficients of the equation will not reveal any information about the data and the adversary will require to know the secret factors of $n$. For this we can use an equation such as the one below:
\begin{equation}\label{eq11}
    x^k+\sum_{i=1}^{k-1}a_{k-i}x^{k-i}+d^y\equiv 0\bmod n
\end{equation}
where $y$ is a secretly chosen exponent and $GCD(y,n)=1$. If the coefficients are chosen such that (\ref{eq11}) has $k$ roots then
\begin{equation}\label{eq12}
    \prod_{i=1}^{k}r_i\equiv d^y\bmod n
\end{equation}
Appropriate coefficients may be chosen in a manner similar to that described in previous sections. It is clear that compromise of all the roots and coefficients will at the most reveal $c=d^y\bmod n$. In order to compute the original value of $d$, the adversary will require the factors of $n$ which are held secret by the sensor which owns the data.

\section{Addressing the Data Partitions}
The previous sections consider the security of the proposed scheme when prime $p$ and composite $n$ are public knowledge. However, there is nothing that compels the user to disclose the values of $p$ and $n$. If we assume that $p$ and $n$ are secret values, then the partitions may be stored in the form of an ``encrypted link list'', which is a list in which every pointer is in encrypted form and in order to find out which node the present node points to, a party needs to decrypt the pointer which can be done only if certain secret information is known.

If we assume that $p$ and $n$ are public values then the pointer can be so encrypted that each decryption either leads to multiple addresses or depends on the knowledge of the factors or both. Only the legitimate party will know which of the multiple addresses is to be picked.

Alternatively, one may use a random number generator and generate a random sequence of sensor IDs using a secret seed. One way to generate this seed may be to find the hash of the original data and use it to seed the generated sequence and keep the hash secret.

\section{Future Work and Other Applications}

Our approach leads to interesting research issues such as the optimal way to distribute the data pieces in a network of $n$ sensors. Also in the case when the sensors are moving, one needs to investigate as to how the partitions need to be reallocated so that the original sensor is always able to access the pieces when needed. Further, questions of load balancing so that no one sensor is storing a very large number of partitions and the partitions are as evenly distributed over the network as possible, needs to be investigated.

Yet another application of the presented scheme is in Internet voting protocols. Internet voting is a challenge for cryptography because of its opposite requirements of confidentiality and verifiability. There is the further restriction of ''fairness" that the intermediate election results must be kept secret. One of the ways to solve this problem is to use multiple layers of encryption such that the decryption key for each layer is available with a different authority. This obviously leaves open the question as to who is to be entrusted the encrypted votes.

A more effective way to implement fairness would be to avoid encryption keys altogether and divide each cast ballot into $k$ or more pieces such that each authority is given one of the   pieces \cite{ref22}. This solves the problem of entrusting any single authority with all the votes and if any of the authorities (less than the threshold) try to cheat by deleting or modifying some of the cast ballots, then the votes may be recreated using the remaining partitions. Such a system implicitly provides a back-up for the votes.

\section{Conclusions}
We have introduced a new distributed data storage scheme for the sensor networks. In this scheme data is partitioned in such a way that each partition is implicitly secure and does not need to be encrypted. Reconstruction of the data requires access to a threshold number of sensors that store the data partition.

An additional variation to the scheme where the data partitions need to be brought together in a definite sequence may be devised. One way to accomplish it is by representing partition in the following manner: \\$p_1, p_1(p_2), p_2(p_3),...$, where $p_i(p_j)$ represents the encryption of $p_j$ by means of $p_i$. Such a scheme will increase the complexity of the brute-force decryption task for an adversary.

%
%


\begin{thebibliography}{5}


\bibitem{ref1} Perrig, A., Szewczyk, R., Wen, V., Culler, D., Tygar, J. : Spins: security protocols for sensor networks. In: Proceedings of ACM Mobile Computing and Networking (Mobicom 01), pp.189--199 (2001)

\bibitem{ref2} Eschenauer, L., Gligor, V. : A Key-management Scheme for Distributed Sensor Networks. In: The 9th ACM Conference on Computer and Communications Security, pp. 41--47 (2002)

\bibitem{ref3} 	Liu, D., Ning, P. : Establishing Pairwise Keys in Distributed Sensor Networks. In: The 10th ACM Conference on Computer and Communcations Security, pp. 52--61 (2003)

\bibitem{ref4} Zhu, S., Setia, S., Jajodia, S. : LEAP: Efficient Security Mechanisms for Large-Scale Distributed Sensor Networks. In: 10th ACM conference on Computer and Communications Security, pp. 62--72 (2003)

\bibitem{ref5} Karlof, C., Wagner, D. : Secure Routing in Wireless Sensor Networks: Attacks and Countermeasures. In: 1st IEEE International Workshop Sensor Network Protocols and Applications, pp. 113--127 (2003)

\bibitem{ref6} Yin, C., Huang, S., Su, P., Gao, C. : Secure routing for large-scale wireless sensor networks. In: Communication Technology Proceedings, vol. 2, pp. 1282--1286 (2003)

\bibitem{ref7} Demirkol, I., Alagoz, F., Delic, H., Ersoy, C. : Wireless Sensor Networks for Intrusion Detection: Packet Traffic Modeling. Communications Letters, IEEE, vol. 10, no. 1, pp. 22--24, (2006)

\bibitem{ref8} Benson, Z., Freiling, F., Cholewinski, P. : Simple Evasive Data Storage in Sensor Networks. In: 17th IASTED International Conference on Parallel and Distributed Computing and Systems: First International Workshop on Distributed Algorithms and Applications for Wireless and Mobile Systems, pp. 779-784 (2005)

\bibitem{ref9} Kak, S. : On the method of puzzles for key distribution. International Journal of Computer and Information Science, vol. 14, pp. 103-109 (1984)

\bibitem{ref10} Kak, S. : Exponentiation modulo a polynomial for data security, International Journal of Computer and Information Science, vol. 13, pp. 337-346 (1983)

\bibitem{ref11} Shamir, A. : How to share a secret, Communication of ACM, vol. 22, no. 11, pp. 612-613 (1979)

\bibitem{ref12} Kak, S. : A cubic public-key transformation. Circuits, Systems and Signal Processing, vol. 26, pp. 353-359 (2007)

\bibitem{ref13} Dickson, L. : Linear Groups with an Exposition of the Galois Field Theory. Dover Publications (1958)

\bibitem{ref14} Rosen, K.H. : Discrete Mathematics and its Applications. McGraw-Hill (2007)

\bibitem{ref15} Greenstein, B., Estrin, D., Govindan, R., Ratnasamy, S., Shenker, S. : DIFS: A Distributed Index for Features in Sensor Networks. Ad Hoc Networks, vol. 1, pp. 333--349 (2003)

\bibitem{ref16} Subramanian, N., Yang, C., Zhang, W. : Securing distributed data storage and retrieval in sensor networks. Pervasive and Mobile Computing, vol. 3, no. 6, pp. 659--676 (2007)

\bibitem{ref17} Wang, G., Zhang, W., Cao, G., La Porta, T. : On supporting distributed collaboration in sensor networks. In: IEEE Military Communications Conference, vol. 2, pp. 752--757 (2003)

\bibitem{ref18} Ye, F., Luo, H., Cheng, J., Lu, S., Zhang, L. : A two-tier data dissemination model for large-scale wireless sensor networks. in: ACM International Conference on Mobile Computing and Networking, pp. 148-159, (2002)

\bibitem{ref19} Zhang, W., Cao, G., La Porta, T. : Data dissemination with ring-based index for sensor networks. In: IEEE International Conference on Network Protocol, pp. 305--314 (2003)

\bibitem{ref20} Renvall,A., Ding, C. : A nonlinear secret sharing scheme. Information Security and Privacy, LNCS vol. 1172, pp. 56-66 (1996)

\bibitem{ref21} Bharucha-Reid, A.T., Sambandham, M. : Random Polynomials. Academic Press, New York (1986)

\bibitem{ref22} Parakh, A., Kak, S. : Internet Voting Protocol Based on Implicit Data Security. In:  Proceedings of 17th International Conference on Computer Communications and Networks, ICCCN '08, pp. 1--4 (2008)



\end{thebibliography}
\end{document}